\documentclass[12pt,preprint]{aastex}
\usepackage{graphicx}

\newcommand{\ms}{\mbox{m\,s$^{-1}~$}}

\newcommand{\mse}{\mbox{m\,s$^{-1}$}}

\newcommand{\mjupe}{$M_{\rm Jup}$}

\newcommand{\feh}{\ensuremath{[\mbox{Fe}/\mbox{H}]}}
\newcommand{\rphk}{\ensuremath{R'_{\mbox{\scriptsize HK}}}}
\newcommand{\lrphk}{\ensuremath{\log{\rphk}}}
\newcommand{\caii}{\ion{Ca}{2} H \& K}

\newcommand{\msini}{\ensuremath{M \sin i}}

\newcommand{\plbPer}{\ensuremath{58.11247 \pm 0.0003}}             
\newcommand{\plbTc}{\ensuremath{15621.637 \pm 0.0156}}      
\newcommand{\plbTp}{\ensuremath{15626.199 \pm 0.024}}      
\newcommand{\plbEcc}{\ensuremath{0.52883 \pm 0.00103}}      
\newcommand{\plbOm}{\ensuremath{172.923 \pm 0.139}}      
\newcommand{\plbK}{\ensuremath{475.133 \pm 0.9102}}      
\newcommand{\plbMsini}{\ensuremath{7.659 \pm 0.0975}}      
\newcommand{\plbA}{\ensuremath{0.2931 \pm 0.00181}}      
\newcommand{\plcPer}{\ensuremath{1749.83 \pm 0.57}}             
\newcommand{\plcTc}{\ensuremath{15599.9 \pm 1.187}}      
\newcommand{\plcTp}{\ensuremath{15521.3 \pm 2.2}}      
\newcommand{\plcEcc}{\ensuremath{0.2113 \pm 0.00171}}      
\newcommand{\plcOm}{\ensuremath{64.87 \pm 0.5113}}      
\newcommand{\plcK}{\ensuremath{297.70 \pm 0.618}}      
\newcommand{\plcMsini}{\ensuremath{17.193 \pm 0.21}}      
\newcommand{\plcA}{\ensuremath{2.8373 \pm 0.018}}      
\newcommand{\plGamma}{\ensuremath{-46.533 \pm 0.552}}      
\newcommand{\plTrend}{\ensuremath{-0.00868 \pm 0.00025}}      
\newcommand{\plNobs}{\ensuremath{130}}      
\newcommand{\plRMS}{\ensuremath{3.90}}      
\newcommand{\plChiNu}{\ensuremath{1.44}}      
\newcommand{\plMv}{\ensuremath{4.198}}      
\newcommand{\plBV}{\ensuremath{0.724}}      
\newcommand{\plVmag}{\ensuremath{6.92}}      
\newcommand{\plDeltamag}{\ensuremath{1.05}}    
\newcommand{\plDist}{\ensuremath{37.4  \pm  1.0}}      
\newcommand{\plFeh}{\ensuremath{+0.04 \pm 0.03}}      
\newcommand{\plTeff}{\ensuremath{5491 \pm  44}}      
\newcommand{\plVsini}{\ensuremath{2.20 \pm 0.50}}      
\newcommand{\plLogg}{\ensuremath{4.07 \pm 0.06}}      
\newcommand{\plMstar}{\ensuremath{0.995 \pm 0.019}}      
\newcommand{\plRstarIso}{\ensuremath{1.51 \pm 0.06}}      
\newcommand{\plLogRphk}{\ensuremath{-5.088}}      
\newcommand{\plSval}{\ensuremath{0.148}}      

\shorttitle{Transit Search for HD~168443b}
\shortauthors{Pilyavsky et al.}
\slugcomment{Submitted for publication in the Astrophysical Journal}

\begin{document}

\title{A Search for the Transit of HD~168443\lowercase{b}: Improved Orbital Parameters and Photometry }

\author{
 Genady Pilyavsky\altaffilmark{1},
 Suvrath Mahadevan\altaffilmark{1,2},
 Stephen R. Kane\altaffilmark{3},
 Andrew W. Howard\altaffilmark{4,5},
 David R. Ciardi\altaffilmark{3},
 Chris de Pree\altaffilmark{6},
 Diana Dragomir\altaffilmark{3,7}, 
 Debra Fischer\altaffilmark{8}, 
 Gregory W. Henry\altaffilmark{9}, 
 Eric L. N. Jensen\altaffilmark{10}, 
 Gregory Laughlin\altaffilmark{11}, 
 Hannah Marlowe\altaffilmark{6}, 
 Markus Rabus\altaffilmark{12},
 Kaspar von Braun\altaffilmark{3},
 Jason T. Wright\altaffilmark{1,2},
 Xuesong X. Wang\altaffilmark{1} 
 }
\email{gcp5017@psu.edu}
\email{suvrath@astro.psu.edu}
\altaffiltext{1}{Department of Astronomy and Astrophysics,
 Pennsylvania State University, 525 Davey Laboratory, University
 Park, PA 16802}
\altaffiltext{2}{Center for Exoplanets \& Habitable Worlds, Pennsylvania State University, 525 Davey Laboratory, University Park, PA 16802}
\altaffiltext{3}{NASA Exoplanet Science Institute, Caltech, MS 100-22,
 770 South Wilson Avenue, Pasadena, CA 91125}
\altaffiltext{4}{Department of Astronomy, University of California,
 Berkeley, CA 94720}
\altaffiltext{5}{Space Sciences Laboratory, University of California,
 Berkeley, CA 94720}
 \altaffiltext{6}{Department of Physics and Astronomy, Agnes Scott College, 141 East College Avenue, Decatur, GA 30030, USA}
\altaffiltext{7}{Department of Physics \& Astronomy, University of
 British Columbia, Vancouver, BC V6T1Z1, Canada}
\altaffiltext{8}{Department of Astronomy, Yale University, New Haven,
 CT 06511}
\altaffiltext{9}{Center of Excellence in Information Systems, Tennessee
 State University, 3500 John A. Merritt Blvd., Box 9501, Nashville,
 TN 37209}
\altaffiltext{10}{Dept of Physics \& Astronomy, Swarthmore College, Swarthmore, PA 19081}
\altaffiltext{11}{UCO/Lick Observatory, University of California, Santa
 Cruz, CA 95064}
\altaffiltext{12}{Departamento de Astonom\'ia y Astrof\'isica,
 Pontificia Universidad Cat\'olica de Chile, Casilla 306, Santiago
 22, Chile}

\begin{abstract}
The discovery of transiting planets around bright stars holds the potential 
to greatly enhance our understanding of planetary atmospheres. In this work 
we present the search for transits of HD168443b, a massive planet orbiting 
the bright star HD~168443 ($V=6.92$) with a period of 58.11 days. The high 
eccentricity of the planetary orbit ($e=0.53$) significantly enhances the 
a-priori transit probability beyond that expected for a circular orbit, 
making HD~168443 a candidate for our ongoing Transit Ephemeris Refinement 
and Monitoring Survey (TERMS). Using additional radial velocities from 
Keck-HIRES, we refined the orbital parameters of this multi-planet system 
and derived a new transit ephemeris for HD168443b. The reduced uncertainties 
in the transit window make a photometric transit search practicable. 
Photometric observations acquired during predicted transit windows were 
obtained on three nights. CTIO 1.0~m photometry acquired on 2010 September 7 
had the required precision to detect a transit but fell just outside of our 
final transit window. Nightly photometry from the T8 0.8~m Automated 
Photometric Telescope (APT) at Fairborn Observatory, acquired over a span 
of 109 nights, demonstrates that HD~168443 is constant on a time scale of
weeks. Higher-cadence photometry on 2011 April 28 and June 25 shows no 
evidence of a transit. We are able to rule out a non-grazing transit of 
HD168443b.
\end{abstract}



\section{Introduction}
\label{introduction}
\indent The number of known exoplanets has grown rapidly in the last decade, 
with over 600 confirmed exoplanets known to date\footnote{exoplanet.eu}. 
While most of these planets have been discovered using radial velocity 
techniques, the number of known transiting planets has increased 
significantly due to dedicated transit surveys like the space-based Kepler 
\citep{Borucki11} and CoRoT \citet{Barge08} missions, and ground-based 
transit searches like the Hungarian Automated Telescope Network (HATNet) 
\citet{Bakos04}, SuperWASP \cite{Pollacco06}, and XO \citep{McCullough05}. 
The price for efficient operation of these wide-field transit surveys, though, 
is that most of the candidate stars tend to be fainter than those being 
surveyed by radial velocity. Of the over one hundred transiting planet 
host stars known, the sample of bright stars ($V<9$) with transiting planets 
is still limited to only nine stars: HD209458 \citep{Charbonneau00, 
Henry00}, HD189733 \citep{Bouchy05}, HD149026 \citep{Sato05}, HD17156 
\citep{Barbieri07}, HD80606 \citep{Moutou09, Fossey09}, HD97658 \citep{Henry11} and 55 Cnc e 
\cite{winn01} (all of which were discovered by radial velocity surveys), 
while WASP-33b and HAT-P-2b were discovered by transit surveys and 
confirmed by radial velocity followup. 

The discovery of additional bright transiting planet hosts is advantageous 
in further enabling studies of the atmospheric constituents of exoplanets. 
Even with the largest ground-based telescopes, transmission spectroscopy 
to probe the atmospheres of these exoplanets has, largely, been accomplished 
only for the brightest targets. Using high-resolution spectroscopy on 8--10~m 
telescopes, \citet{Redfield08} and \citet{Snellen08} detected sodium 
absorption in the transmission spectra of HD189733b and HD209458b, 
respectively. More recently, using the narrowband tunable filter imager 
on the 10~m GTC, \citet{Colon10} and \citet{Sing11} have detected the 
signature of potassium absorption in the atmospheres of HD~80606b and XO-2b.

Our ongoing Transit Ephemeris Refinement and Monitoring Survey (TERMS, 
\citet{Kane09}) project focuses on bright stars ($V<9$) with known exoplanets 
and orbital periods greater than 10 days in an effort to refine the orbital 
parameters with additional radial velocity observations and then observe the 
targets photometrically within their revised transit windows. Transits 
detected around such bright stars would provide perfect candidates for 
spectroscopic follow up. In addition, with periods greater than ten 
days, the planet population searched by TERMS is not easily duplicable by 
ongoing ground-based transit surveys as demonstrated by \citet{vonBraun09}. 
In \citet{Kane11} we presented the ephemeris revision and the search for a 
transit around HD~156846. In this paper, we present additional radial 
velocities and refine the transit ephemeris for the bright star HD~168443, 
which is known to have multiple companions. We present new photometry that
allows us to rule out transits of HD~168443b. 


\section{HD~168443}
\label{motivation}
HD~168443 (GJ~4052, HIP~89844, TYC~5681-1576-1) is a bright ($V=6.92$) 
G5 dwarf known to possess two substellar companions, forming a dynamically 
active system \citep{Veras07}. HD~168443b \citep{mar99,wright2009b} has a 
reported $M_p$sin$i=7.8\pm 0.259 M_{Jup}$, an orbit with a period of 
58.11 days and a large eccentricity of $e=0.53$. HD~168443c, a brown dwarf 
companion \citep{Udry02, wright2009b}, has $M_p$sin$i=17.5\pm 0.65 M_{Jup}$, 
an orbital period of $\sim$1748 days and a moderate eccentricity $e=0.21$. 
Using the \citet{van07} re-reduction of the $Hipparcos$ data, \citet{Reffert11} 
derive a mass of $30.3^{+9.4}_{-12.2}M_{Jup}$ and a $3\sigma$ upper mass limit 
of $65M_{Jup}$, confirming that this object is indeed substellar. The 
$3\sigma$ lower limit does not exclude an inclination of 90$^\circ$, so 
the minimum mass derived from the radial velocities applies. From CORALIE 
radial velocities and the \citet{van07} $Hipparcos$ re-reductions, 
\citet{Sahlmann11} concluded that, while their formal solution for the mass 
of HD~168443c matched that of \citet{Reffert06}, the mass was of "low 
confidence." They are unable to set an upper mass limit because the 
radial velocity orbit is not fully covered by their CORALIE obsevations. 
Dynamical simulations by \citet{Veras11} show that almost no stable systems 
can exist for mutual inclinations between HD~168443b and c of 
60$^\circ$--120$^\circ$.

The eccentric orbit of HD~168443b increases its transit probability 
significantly above what one would expect for a planet in a circular orbit 
with the same period. The new orbital parameters, along with the formalism 
outlined in \citet{Kane09}, result in a transit probability of 3.7\% compared 
to 2.5\% for a circular orbit. While the atmospheric scale heights for 
massive, relatively cold planets are expected to be small \citep{Madjar11}, 
the brightness of HD~168443 (more than twice as bright as HD~209458, the 
third brightest star known to have a transiting planet) makes such 
detections possible with large ground-based telescopes. 

The discovery of transits of the inner planet would also constrain the 
possible inclinations of the outer brown dwarf companion, enabling 
additional dynamical investigations. The predicted transit probability 
of 3.7\%, coupled with the fact that HD~168443 is a very bright star, 
makes it an intriguing target in our ongoing attempts to discover 
long-period transiting planets.


\section{Stellar Properties}
\label{steprop}

We used Spectroscopy Made Easy \citep[SME;][]{val96} to fit high-resolution
Keck-HIRES spectra of HD~168443, applying the wavelength intervals, line 
data, and methodology of \citet{val05}. We further constrained the surface 
gravity using Yonsei-Yale (Y$^2$) stellar structure models \citep{dem04} 
and revised \textit{$Hipparcos$} parallaxes \citep{van07}, with the iterative
method of \citet{val09}. The resulting stellar parameters listed in
Table \ref{tab:stellar_params} are effective temperature, surface
gravity, iron abundance, projected rotational velocity, mass, and
radius. HD~168443 lies \plDeltamag\ mag above the $Hipparcos$ average
main sequence ($M_V$ versus $B-V$) as defined by \citet{wri05}. These
properties are consistent with an evolved metal-rich G5 star. The stellar 
radius, $R_{\star} = \plRstarIso\ R_{\sun}$, is crucial for estimating the 
depth and duration of a planetary transit. This value is consistent with 
the interferometrically measured $R_{\star} = 1.58 \pm 0.06\ R_{\sun}$ 
\citep{vanBelle09}.

In addition, we computed the level of stellar activity in HD~168443 from 
the strength of the \caii\ lines, which give calibrated $S_\mathrm{HK}$ 
values on the Mt.\ Wilson scale and \lrphk\ values \citep{isa10}. 
The median of \lrphk\ and $S_\mathrm{HK}$ values are listed in 
Table \ref{tab:stellar_params} and demonstrate that HD~168443 is 
chromospherically quiet. Additional examination of the available history 
of \caii\ measurements show no significant long-term variation in 
$S_\mathrm{HK}$. 

\begin{center}
\begin{deluxetable}{lcr}
 \tabletypesize{\footnotesize}
 \tablecaption{Stellar Properties
 \label{tab:stellar_params}}
 \tablewidth{0pt}
 \tablehead{
 \colhead{Parameter} & 
 \colhead{Value} &
 \colhead{Reference}
 }
 \startdata
 $M_V$ & \plMv & \citet{van07} \\
 $B-V$ & \plBV & \citet{Bessell01} \\
 $V$ & \plVmag & \citet{Bessell01} \\
 Distance (pc) & \plDist & \citet{van07} \\
 $T_\mathrm{eff}$ (K) & \plTeff & This work \\
 log\,$g$ & \plLogg & This work \\
 \feh & \plFeh & This work \\
 $v$\,sin\,$i$ (km\,s$^{-1}$) & \plVsini & This work \\
 $M_{\star}$ ($M_{\sun}$) & \plMstar & This work \\
 $R_{\star}$ ($R_{\sun}$) & \plRstarIso & This work \\
 \lrphk & \plLogRphk & This work \\
 $S_\mathrm{HK}$ & \plSval & This work
 \enddata
\end{deluxetable}
\end{center}


\section{Keck-HIRES RV Measurements and Revised Orbital Model}
\label{sec:rv}

\subsection{Measurements}

We observed HD~168443 using the standard procedures of the California 
Planet Search (CPS) for the HIRES echelle spectrometer \citep{vog94} on 
the 10~m Keck I telescope. These measurements span fifteen years, 1996 July 
to 2011 March, and comprise one of the longest RV datasets presented for a 
star with one or more known planets. The initial measurements in this 
time series were used to discover the two planets \citep{mar99,mar01} while 
later measurements refined the orbits \citep{wright2009b}. The full set of 
measurements presented here refines the orbit further and gives an 
accurate predicted transit ephemeris with which we search for photometric 
transits.

The \plNobs\ Keck RV measurements (Table \ref{tab:keck_rvs}) were made 
from observations with an iodine cell mounted directly in front of the
spectrometer entrance slit. The dense set of molecular absorption lines 
imprinted on the stellar spectra provide a robust wavelength
fiducial against which Doppler shifts are measured, as well as strong
constraints on the shape of the spectrometer instrumental profile at
the time of each observation \citep{mar92,val95}. We measured
the Doppler shift of each star-times-iodine spectrum using a modeling
procedure descended from \citet{but96} as described in \citet{how09}.
The times of observation (in barycentric Julian days), relative RVs,
and associated errors (excluding jitter) are listed in Table
\ref{tab:keck_rvs}. We also observed HD~168443 with the iodine cell removed
to construct a stellar template spectrum for Doppler modelling and to
derive stellar properties.

\subsection{Keplerian Model}

We modelled the Keck RVs as the superposition of the Keplerian interactions 
from two planets with the star, plus a linear trend in velocity due to a 
distant and massive third companion. We used the orbit fitting techniques 
described in \citet{how10} and the partially linearized, least-squares 
fitting procedure described in \citet{wri09}. Each velocity measurement 
was assigned a weight, $w$, constructed from the quadrature sum of the 
measurement uncertainty ($\sigma_{\mathrm{RV}}$) and a jitter term 
($\sigma_{\mathrm{jitter}}$), i.e.,
\ $w$ = 1/($\sigma_{\mathrm{RV}}^2+\sigma_{\mathrm{jitter}}^2$). 
We chose jitter values of $\sigma_{\mathrm{jitter}} = 3$ and 2~\ms for 
measurements before and after the HIRES upgrade in August 2004.
These values are consistent with the expected jitter of a slightly evolved 
early G star observed with Keck/HIRES \cite{wri05}. Possible sources of 
jitter include stellar pulsation, magnetic cycles, granulation, undetected 
planets, and instrumental effects \citep{isa10,wri05}.

\begin{figure}[ht]
\includegraphics[width=0.8\textwidth]{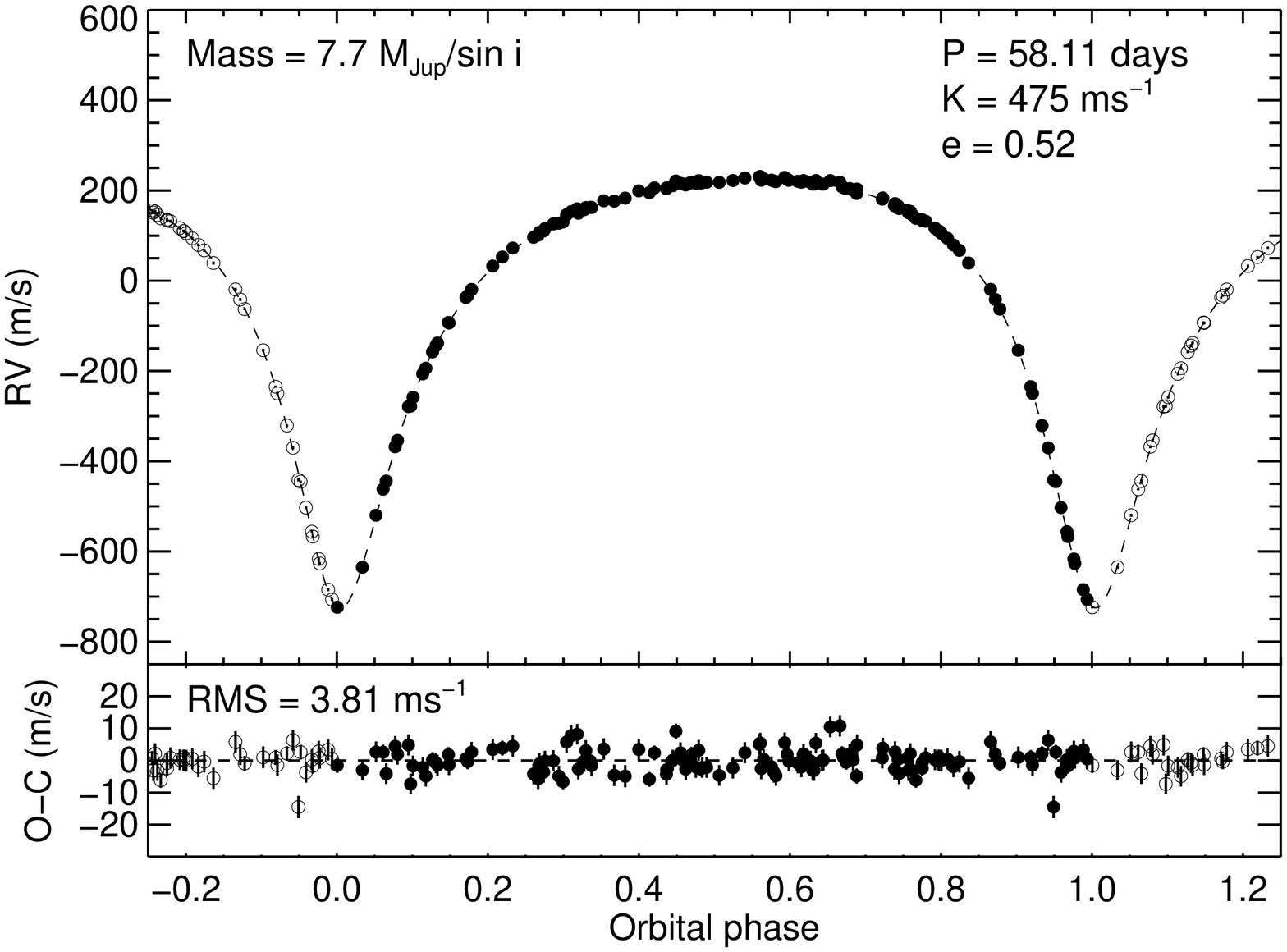}
\includegraphics[width=0.8\textwidth]{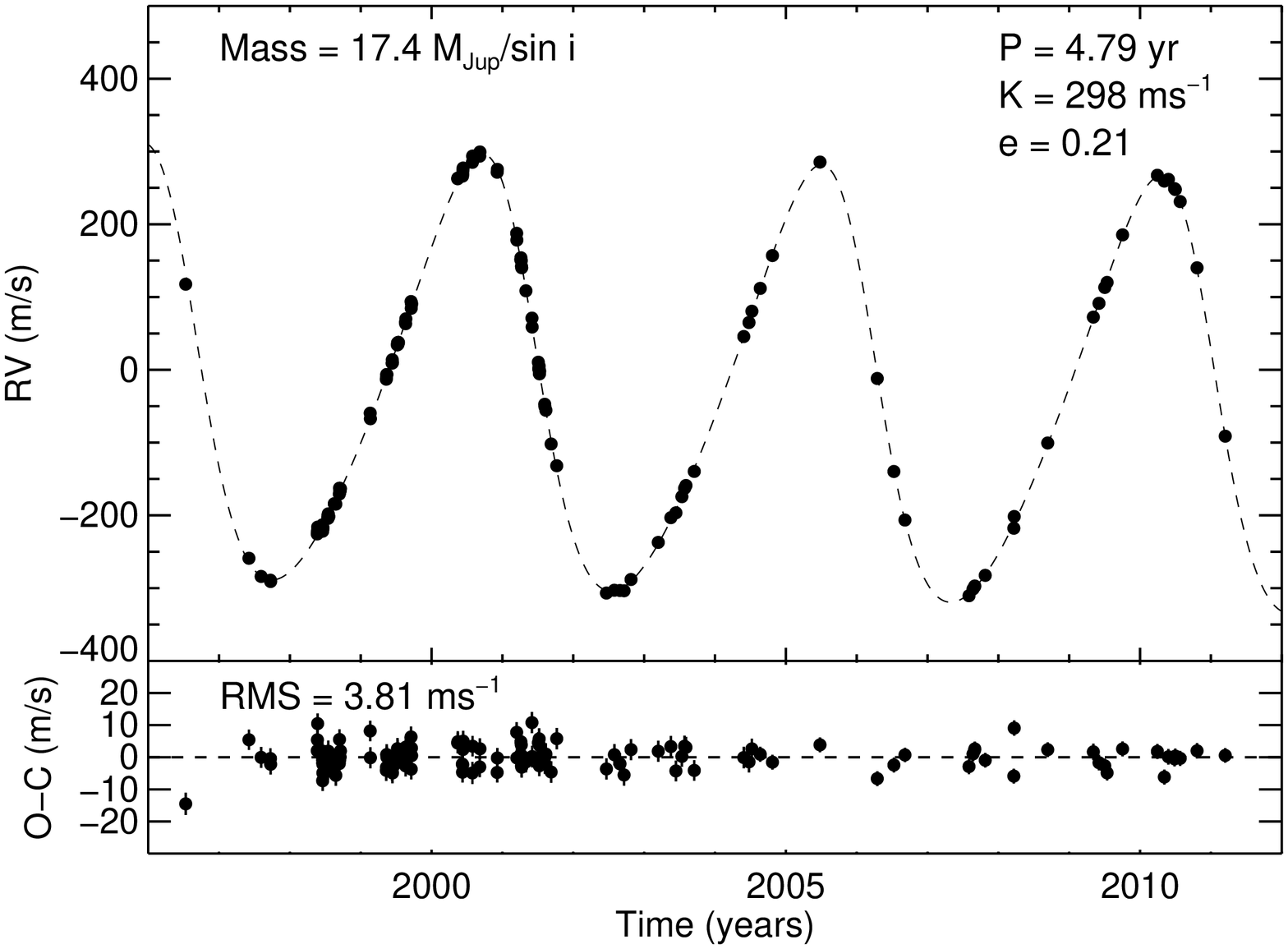} 
\caption{RV measurements of HD~168443 from Keck-HIRES (filled circles) 
with Keplerian orbital model (dashed lines). The top panel shows the RVs 
phased to the orbital period of HD~168443~b with the model for the other 
planet and linear trend subtracted. Open circles represent the same RV 
measurements wrapped one orbital phase. The bottom panel shows the RV time 
series illustrating the variations due to HD~168443~c, with the linear 
velocity trend and the orbit of HD~168443~b subtracted.
}
\label{fig:rvmodel}
\end{figure}

Our best-fit orbital model is presented in Table \ref{tab:orbital_models} and 
Figure \ref{fig:rvmodel}. The Keplerian parameter uncertainties for each 
planet were derived using a Monte Carlo method \citep{mar05} and account 
for correlations between parameter errors. Uncertainties in \msini\ and 
$a$ reflect uncertainties in $M_{\star}$ and the orbital parameters. 
We considered and rejected more complicated models having a third planet 
and/or a quadratic velocity trend because of statistically insignificant 
changes in $\chi^2$ compared to the adopted model 
(Table \ref{tab:orbital_models}).


\section{Transit Ephemeris Refinement}
\label{ephemeris}

\begin{figure}[h]
\includegraphics[angle=270,width=8.2cm]{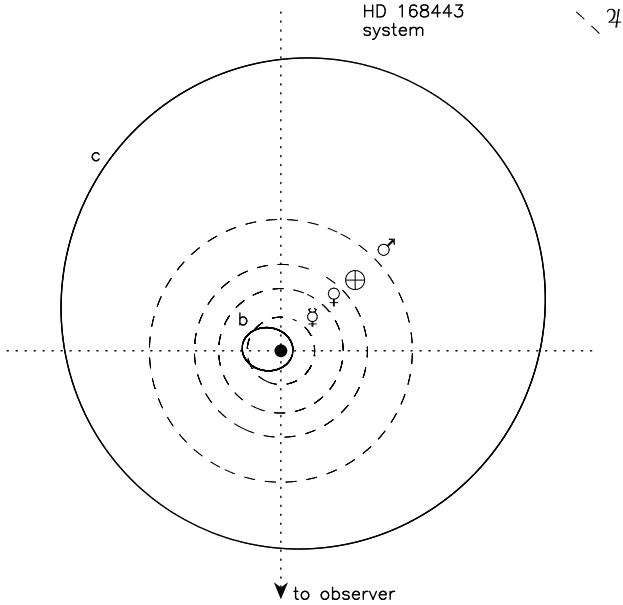}
\caption{The orbits of the planets HD~168443b and HD~168443c shown in solid 
lines. Orbits of Mercury (eccentricity set to zero for clarity) through 
Jupiter are represented with dashed ovals.}
\label{fig:orbit}
\end{figure}

The revised orbital solution presented in Table \ref{tab:orbital_models} for 
this multi-planet system, along with the stellar properties in Table 
\ref{tab:stellar_params}, allow us to construct an accurate transit 
ephemeris from which to conduct a search for transits. As shown by 
\citet{kan08}, the transit probability of a planet is
intricately related to both the orbital eccentricity and the argument of
periastron. Figure \ref{fig:orbit} depicts the orbits of the planets 
relative to the observer line-of-sight and shows how the eccentricities 
and orientations affects the star-planet separation along that line.


We use the models of \citet{bod03} to estimate a radius for HD~168443b of 
$R_p = 1.11 R_{Jup}$, which takes into account orbital parameters and the 
stellar flux received by the planet. This results in a transit probability 
of 3.7\% and a predicted transit depth of 0.6\%. The predicted transit 
duration is 0.36~days, or 8 hours and 40 minutes. The time of mid transit 
shown in Table \ref{tab:orbital_models} is for 2011 March 1. The 
calculation was performed using a Monte-Carlo bootstrap which propagates 
the uncertainties in the orbital parameters forward to the time of transit. 
The uncertainty in this predicted time is small; only $\sim 35$ minutes. 
Thus, the predicted duration of the transit window for this date is 
9 hours and 50 minutes (the sum of the predicted duration $\pm$ 1-$\sigma$ 
deviation), overwhelmingly dominated by the predicted transit duration 
rather than the uncertainty associated with the orbital parameters. The 
predicted duration makes complete coverage of the transit window from a 
single ground-based longitude difficult, though the predicted depth is 
sufficient to rule out transits from observations during times of ingress 
or egress only.


\section{$Hipparcos$ Photometry of HD~168443}
\begin{figure}[h]
\includegraphics[angle=90,width=8.2cm]{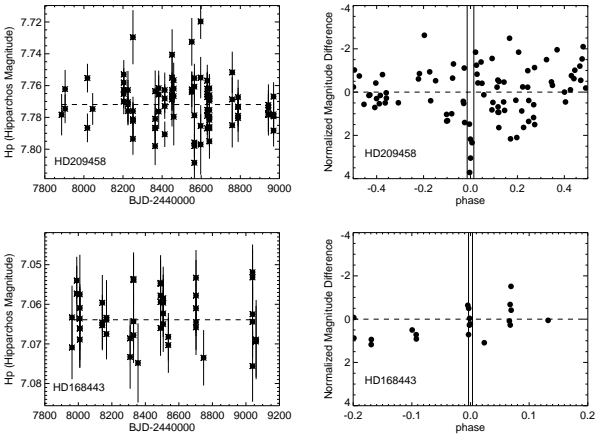}
\caption{$Hipparcos$ Photometry. Top left: Photometric measurements of 
HD~209458 over the course of three years. Top right: the same data plotted 
vs. orbital phase, with the transit corresponding to zero phase. The transit 
is clearly visible and is marked by two vertical lines. Bottom left: 
$Hipparcos$ Photometric measurements of HD~168443 taken over the course of 
three years. Bottom right: Similar to the plot for HD~209458 (following the 
method of \citet{Robichon00}). The predicted window is bounded by two 
vertical lines, with three $Hipparcos$ measurements obtained during the 
predicted transit. No transit is evident.}
 \label{fig:hip}
\end{figure}

The $Hipparcos$ mission observed the brightest stars in the sky over many 
epochs. \citet{Robichon00} and \citet{Castellano00} detected the transit of 
HD~209458b in the $Hipparcos$ epoch photometry, and \citet{Hebrard06} were 
able to detect multiple transits of HD~189733b, leading to a significant 
improvement in the determination of its period. $Hipparcos$ photometry has 
therefore been demonstrated to be precise enough to detect the transit of 
a Hot Jupiter around stars that are fainter than HD~168443 (V=6.92). 
The top panels of Figure \ref{fig:hip} show the $Hipparcos$ photometry of 
HD~209458 plotted against Julian Date and also against orbital phase of the
star's Hot-Jupiter companion. A few observations fall within the modern 
transit window and show a clear dimming. The bottom two panels are similar 
plots for HD~168443. Using our new radial velocity observations, we 
followed the prescription of \citet{Robichon00} to create the phased 
plot with the predicted transit set at zero phase. Three $Hipparcos$ 
photometric observations acquired around BJD 2448050 lie inside of our 
predicted transit window. Unlike the case for HD~209458, the $Hipparcos$ 
photometry show no evidence for a transit. However, the expected transit 
depth for HD~168443b is significantly smaller than HD~209458; the $Hipparcos$ 
photometry alone would have been an unreliable guide to prove or disprove 
the occurrence of transits in HD~168443. Nevertheless, we advocate such a 
check for bright stars with sufficiently precise orbital ephemerides. 


\section{Photometry of the Revised Transit Window}
In addition to the new TERMS photometry presented below, we also observed 
HD~168443 with the Southeastern Association for Research in Astronomy (SARA) 
0.6~m telescope at Cerro Tololo Inter-American Observatory. However, due 
to poor weather conditions, the SARA-S measurements exhibited scatter that 
was significantly higher than the predicted transit depth. Therefore, we 
do not discuss these data further, but we note that planning such multi-site 
observations is necessary in our TERMS search for long-period transits.

\subsection{Cerro Tololo Inter-American Observatory (CTIO)}

Before we had access to the latest Keck-HIRES radial velocities, we used 
published orbital parameters \citep{wright2009b} to calculate the transit 
ephemeris and to schedule an observing run for 2010 September 7with the 
CTIO 1.0~m telescope and Y4KCam CCD detector. The observations were made 
through a Johnson-Morgan R-band filter; instrumental magnitudes of HD~168443 
and comparison stars were extracted from the images with an IDL 
implementation of DAOPHOT \citep{stat01}. Relative fluxes (Everett \& Howell 
2001) of HD~168443 were computed with respect to the two stable comparison 
stars (TYC~5681-1450-1 and TYC~5681-1458-1).

The results are plotted in Figure 4. The solid curve represents the predicted 
transit fluxes computed from our new orbital elements in Table~2, assuming 
on-time, central transits with a predicted depth of 0.6\% or 0.006 flux units. 
The scatter of the measurements is 0.0027, easily sufficient to detect the 
predicted transits. However, these measurements cover only the later part
of the predicted transit window; an egress late by 1$\sigma$ and 3 $\sigma$ 
are represented with dashed blue and red lines, respectively. The CTIO data
show only that late transits do not occur.

We have presented these measurements because CTIO photometric monitoring is 
an essential component of the TERMS strategy. Had the ephemerides (based on 
orbital parameters in the literature) been more precise, the CTIO photometry 
has the requisite precision to detect a predicted on-time transit. Our 
results demonstrate the precision achievable from CTIO with a typical TERMS 
target.


\subsection{Fairborn Observatory}
We obtained additional photometric observations of HD~168443 with the T8 
0.8~m automatic photometric telescope (APT) at Fairborn Observatory in 
southern Arizona. The T8 APT uses a two-channel precision photometer with 
two EMI 9124QB bi-alkali photomultiplier tubes to make simultaneous 
measurements in the Str\"omgren $b$ and $y$ passbands. The telescope was 
programmed to make nightly differential brightness measurements of HD~168443 
with respect to the comparison star HD~166664. Three consecutive differential 
measurements were co-added to create a single nightly differential magnitude. 
To improve the precision of these brightness measurements, we combined the 
individual $b$ and $y$ differential magnitudes into a mean $(b+y)/2$ 
"passband". The typical precision of a single observation on good nights is 
$\sim~0.0015$ mag. See \citep{Henry99} for further details on telescope 
design and operations, data reduction, calibrations, and data precision.


Between 2011 March 2 and June 19, the APT collected 107 nightly observations 
of HD~168443 with respect to HD~166664; the differential $(b+y)/2$ 
magnitudes were converted to relative fluxes and are plotted in Figure~5. 
The nightly observations scatter about their mean flux, indicated by the 
dashed line in Figure~5, with a standard deviation of 0.0016, which is 
consistent with constant stars. Periodogram analyses from one to 100 days 
reveal no significant periodicity. We conclude that HD~168443 is constant 
on its rotation timescale. A least-squares sine fit of the nightly 
observations on the orbital period of 168443b gives a semi-amplitude of 
0.00026 $\pm$ 0.00023 flux units. This very low limit to brightness 
variability in HD~168443 on the 58-day orbital period indicates that 
rotational modulation of starspots is not the cause of the radial velocity 
variations, see e.g., \citep{Queloz+2001}.

We also used the T8 APT to monitor HD~168443 on two nights in the 2011 
observing season when our new ephemeris predicted additional transit events. 
Each monitoring observation consists of a single differential measurement
rather than the mean of three observations, as we used for the nightly 
observations; thus the monitoring observations will have more scatter than 
the nightly observations. 


On 2011 April 28 UT, an observable egress was to occur at BJD~2,455,679.937. 
We successfully monitored the star and obtained the 80 measurements plotted 
in Figure~6. The star was still two months before its opposition, so the 
start time was delayed until the star rose above an airmass of $\sim 2.0$; 
the observations ended 3.3 hours later (at dawn). The sudden increase of 
the scatter after predicted egress is the result of a plume of smoke from 
one of the many wildfires burning throughout southern Arizona at the time. 
The standard deviation of the entire data set is 0.0040 flux units. The 
solid curve in Figure~6 represents the predicted fluxes for an on-time, 
central transit. The dotted blue lines represent the $\pm~1\sigma$ 
uncertainty in the predicted time of central transit. The difference between 
the mean flux of the 47 pre-egress observations and the 33 post-egress 
observations is only $0.0005\pm0.0011$. Given the tight limit on the transit 
time (35 min or 0.024 day), these observations rule out central transits 
with the predicted depth of 0.006 with a SNR of $\sim~5:1$.


The next predicted transit was calculated to occur 58 days later on 
2011 June 25 UT, centered at BJD~2,455,737.862. HD~168443 was at opposition 
during this transit, so we were able to acquire 165 observations over an 
interval of 7.1 hours. The declination of HD~168443 is approximately 
$-10\arcdeg$, so the airmass values at the start and at the end of the night 
were 2.25 and 2.51, respectively, for the the east and west observing limits. 
Atmospheric extinction was also significantly higher than normal at the 
beginning of the night due to airborne dust, which gradually settled out
over the course of the night. Therefore, in addition to reducing the transit 
observations with larger than normal extinction coefficients, we also removed 
a linear trend of approximately 0.008 flux units via a linear least-squares 
fit. The residuals from the line fit are plotted in Figure~7, again 
compared with predicted transit fluxes. The overall scatter of the 165 
observations is 0.0039 flux units. 

Finally, we estimate an upper limit to the possible transits of HD~168443b. 
The mean {\it differential magnitudes} of the T8 nightly observations in 
Figure~5 and of the mean of the first and second T8 transit monitoring 
observations in Figures 6 \& 7, before they were converted to relative 
fluxes, are -0.24041 $\pm$ 0.00017 mag, -0.23961 $\pm$ 0.00050 mag, and 
-0.24044 $\pm$ 0.00037 mag, respectively. So, the observed difference 
between the mean of the nightly observations and the mean of the transit 
observations is only 0.00038 $\pm$ 0.00045 mag or 0.00035 $\pm$ 0.00041 
flux units. Therefore, we should have been able to detect a transit depth 
of 0.2\% or 0.002 at a SNR of almost 5:1, but our observations show no 
evidence of a transit.

\subsection{Full Coverage of the Transit Window}


The entire transit window was covered during the three monitoring nights. 
Initially, we were able to rule out only a late egress using 1-meter CTIO 
photometry from 2010 September 8 with (3-$\sigma$) confidence. We further 
confirmed this result with the APT photometry on June 25 by ruling out a late 
ingress (3-$\sigma$). The early (1-$\sigma$) and on time egress was ruled out 
with APT photometry from April 28, 201. In Figure \ref{fig:APTPHOT_phase} 
we present a phase plot containing all thee monitoring nights. Since both 
data sets from the T8 APT used the same reference star, we placed the median 
of the measurements on the same y-axis scale. The CTIO data set was adjusted 
with an offset using the overlapping points from CTIO and the APT. The 
center of predicted transit is at phase 0.0, marked by a solid vertical 
line. The dotted blue lines represent 1-$\sigma$ early and late windows while 
the dashed red lines represent 3-$\sigma$ deviation from the center of the 
predicted transit.


\section{Discussion}
As part of our ongoing Transit Ephemeris Refinement and Monitoring Survey 
(TERMS), we present revised orbital parameters for the HD~168443 system, 
based on 130 radial velocity measurements with Keck-HIRES that span almost 
15 years. Using the transit ephemerides derived from the revised orbital 
parameters, we searched for a transit using telescopes from CTIO, Fairborn 
Observatory, and SARA-S. We find no evidence of a detectable transit. The 
presence of a non-grazing transit corresponding to our model (1.1$R_J$ planet) 
is ruled out at a high level of confidence with high precision photometry 
acquired by the APT. Grazing transits or transits with a planet radii as 
small as 0.58 R$_{Jup}$ (which would yield densities much too high) are 
formally excluded at 1-$\sigma$ confidence, though the smaller duration of 
such transits implies that the time of our photometric observations could 
have missed the ingress and egress. Using our orbital solution, and the 
planetary and stellar radii presented in this paper, we derive an upper 
limit on inclination of the system at 87.8 $^\circ$ using methods described 
by \citet{kan08}. 

Even with a number of RV observations, determining precise orbital parameters 
for one component of a multi-planet system can be difficult. The recent 
discovery of transits of 55~Cnc~e by \citet{winn01}, based on a new orbital 
period by \citet{dawson01}, illustrates the insidious impact of aliases and 
harmonics. Nevertheless, these effects can be mitigated by careful 
observation and analysis, and the value of additional precision radial 
velocities and revised orbital parameters cannot be overstated.

While we do not see transits in the HD~168443 system, the experimental 
approach outlined here, combining high precision radial velocity with 
multiple photometric telescope facilities, is worth pursuing in the quest 
for new transiting planets around bright nearby stars.


\section*{Acknowledgements}

This work made use of the SIMBAD database (operated at CDS, Strasbourg, 
France), NASA's Astrophysics Data System Bibliographic Services, and the 
NASA Star and Exoplanet Database (NStED). This work was partially supported 
by funding from the Center for Exoplanets and Habitable Worlds, supported 
by the Pennsylvania State University, the Eberly College of Science, and the
Pennsylvania Space Grant Consortium. The authors would also like to thank 
Andr\'es Jord\'an for providing support for the observations at CTIO. G.W.H. acknowledges support from NASA, NSF, Tennessee State University, and 
the Tennessee Centers of Excellence Program. ELNJ acknowledges support from 
NSF grant AST-0721386. M.R. acknowledges support from ALMA-CONICYT projects 
31090015 and 31080021. We would also like to thank the referee for insightful 
comments that helped us to improve this paper. Finally, the authors wish to 
extend special thanks to those of Hawai`ian ancestry on whose sacred mountain 
of Mauna Kea we are privileged to be guests. Without their generous 
hospitality, the Keck observations presented herein would not have been 
possible.


\begin{deluxetable}{ccc}
 \tabletypesize{\footnotesize}
 \tablecaption{Keck Radial Velocities
 \label{tab:keck_rvs}}
 \tablewidth{0pt}
 \tablehead{
 \colhead{} & \colhead{Radial Velocity} & \colhead{Uncertainty} \\
 \colhead{BJD -- 2440000} & \colhead{(\mse)} & \colhead{(\mse)} 
}
 \startdata
  10276.90890 & -309.29 &    1.73  \\
 10603.01184 &  -33.47 &    1.02  \\
 10665.86781 &  -69.33 &    1.24  \\
 10713.73770 &  -73.02 &    1.14  \\
 10714.76649 &  -71.38 &    1.13  \\
 10955.01039 &   -5.57 &    1.12  \\
 10955.95862 &   -5.13 &    1.02  \\
 10957.07105 &   -4.68 &    1.05  \\
 10981.88012 & -568.69 &    1.01  \\
 10982.89132 & -492.38 &    1.10  \\
 10983.07690 & -472.56 &    1.10  \\
 10983.82231 & -420.40 &    1.16  \\
 10984.06138 & -407.63 &    1.14  \\
 11009.87009 &   22.04 &    1.34  \\
 11010.05994 &   24.94 &    0.93  \\
 11010.85123 &   20.67 &    1.17  \\
 11011.86077 &   25.19 &    1.37  \\
 11012.95413 &   19.19 &    1.14  \\
 11013.06816 &   20.31 &    0.82  \\
 11013.82791 &   15.69 &    1.15  \\
 11013.92983 &   15.33 &    1.20  \\
 11042.95557 & -326.88 &    1.09  \\
 11043.95602 & -276.32 &    1.19  \\
 11050.81406 &  -77.68 &    1.26  \\
 11068.77042 &   54.21 &    1.08  \\
 11069.78596 &   60.86 &    1.09  \\
 11070.79807 &   52.81 &    1.09  \\
 11071.76998 &   50.98 &    1.11  \\
 11072.76271 &   47.77 &    1.26  \\
 11074.78514 &   39.07 &    1.09  \\
 11228.16111 &   91.81 &    1.07  \\
 11229.14942 &   95.62 &    1.32  \\
 11311.04174 &  151.35 &    1.27  \\
 11312.07757 &  137.66 &    1.23  \\
 11313.07589 &  124.69 &    1.18  \\
 11314.08699 &  103.43 &    1.20  \\
 11341.02661 &  109.66 &    1.11  \\
 11341.90588 &  129.71 &    1.03  \\
 11342.97144 &  142.29 &    1.20  \\
 11367.86065 &  214.31 &    1.30  \\
 11368.84314 &  205.20 &    1.40  \\
 11370.01150 &  188.83 &    1.14  \\
 11370.91933 &  171.14 &    0.73  \\
 11371.91024 &  153.12 &    1.11  \\
 11372.87811 &  131.18 &    1.16  \\
 11373.79496 &  105.31 &    1.44  \\
 11409.85005 &  276.28 &    1.14  \\
 11410.84584 &  278.53 &    1.29  \\
 11411.84641 &  288.70 &    1.21  \\
 11438.73808 & -282.82 &    1.19  \\
 11439.73057 & -414.58 &    1.18  \\
 11440.71830 & -527.69 &    1.22  \\
 11441.73997 & -616.71 &    1.20  \\
 11679.04801 & -109.80 &    1.09  \\
 11680.07181 &  -20.27 &    1.29  \\
 11703.01759 &  488.68 &    1.19  \\
 11703.98840 &  488.90 &    1.17  \\
 11705.03899 &  493.45 &    1.27  \\
 11705.95846 &  499.46 &    1.30  \\
 11707.08125 &  502.77 &    1.20  \\
 11754.85883 &  473.14 &    1.66  \\
 11755.90986 &  489.51 &    1.06  \\
 11792.74760 & -338.38 &    1.01  \\
 11793.80008 & -223.10 &    1.25  \\
 11882.68333 &  496.02 &    0.88  \\
 11883.68272 &  498.30 &    0.86  \\
 11983.15885 &  333.08 &    1.01  \\
 11984.15396 &  335.98 &    1.35  \\
 12004.12317 &  357.01 &    1.33  \\
 12005.14737 &  352.21 &    1.37  \\
 12007.13381 &  329.80 &    1.16  \\
 12008.03597 &  311.14 &    1.37  \\
 12009.10491 &  294.16 &    1.24  \\
 12030.98196 &  -27.97 &    1.26  \\
 12061.94969 &  278.32 &    1.37  \\
 12062.96034 &  259.31 &    1.54  \\
 12094.88634 &   78.37 &    1.40  \\
 12096.93360 &  110.11 &    1.52  \\
 12098.01419 &  126.96 &    1.38  \\
 12099.02269 &  145.64 &    1.34  \\
 12099.93319 &  147.47 &    1.23  \\
 12100.94388 &  158.52 &    1.41  \\
 12101.88510 &  171.72 &    1.42  \\
 12127.86347 &   57.89 &    1.51  \\
 12128.79403 &   30.56 &    1.23  \\
 12133.78671 & -210.61 &    1.53  \\
 12160.80699 &   79.11 &    1.48  \\
 12189.76875 & -156.78 &    1.43  \\
 12445.93556 & -192.60 &    1.38  \\
 12486.81165 & -930.27 &    1.38  \\
 12515.75755 &  -85.29 &    1.44  \\
 12536.74799 & -258.80 &    1.39  \\
 12572.69122 &  -74.58 &    1.31  \\
 12713.14689 & -280.58 &    1.33  \\
 12778.02605 & -891.12 &    1.36  \\
 12804.06082 &   12.02 &    1.25  \\
 12834.88152 & -730.70 &    1.31  \\
 12848.80396 & -133.69 &    1.29  \\
 12855.96590 &    0.22 &    1.28  \\
 12898.71740 & -579.54 &    1.16  \\
 13154.04176 &  260.17 &    1.49  \\
 13180.90222 & -183.23 &    1.37  \\
 13195.84965 &   58.60 &    1.13  \\
 13238.88150 & -123.80 &    1.17  \\
 13301.73743 & -565.16 &    1.16  \\
 13546.89064 &  334.18 &    1.14  \\
 13842.12630 &  125.12 &    1.18  \\
 13927.87788 &   -3.76 &    1.15  \\
 13984.83683 &  -51.21 &    0.96  \\
 14314.99539 & -101.88 &    1.26  \\
 14335.95562 & -191.79 &    1.23  \\
 14343.88345 & -620.98 &    1.00  \\
 14344.94159 & -744.93 &    1.11  \\
 14398.74642 & -344.38 &    1.08  \\
 14546.11545 &  -16.53 &    1.17  \\
 14548.15404 &   10.39 &    1.27  \\
 14720.84072 &  102.86 &    1.11  \\
 14956.12833 &  289.18 &    1.53  \\
 14985.11014 & -473.97 &    1.33  \\
 15014.97707 &  332.37 &    1.13  \\
 15026.96186 &  318.51 &    1.22  \\
 15106.74809 & -279.16 &    1.20  \\
 15286.11499 &  173.04 &    1.21  \\
 15322.08481 &  403.90 &    1.21  \\
 15343.00253 &  103.85 &    1.36  \\
 15374.84199 &  453.14 &    1.16  \\
 15378.82083 &  413.14 &    1.21  \\
 15403.81739 &  198.38 &    1.22  \\
 15490.73506 &  346.47 &    1.30  \\
 15636.13540 & -128.87 &    1.05  \\

 \enddata
\end{deluxetable}

\begin{deluxetable}{lc}
 \tablecaption{Keplerian Orbital Model
 \label{tab:orbital_models}}
 \tablewidth{0pt}
 \tablehead{
 \colhead{Parameter} & \colhead{Value} 
}
 \startdata
\noalign{\vskip -3pt}
\sidehead{HD 168443 b}
~~~~$P$ (days) & \plbPer \\
~~~~$T_c\,^{a}$ (JD -- 2,440,000) & \plbTc \\
~~~~$T_p\,^{b}$ (JD -- 2,440,000) & \plbTp \\
~~~~$e$ & \plbEcc \\
~~~~$K$ (m\,s$^{-1}$) & \plbK \\
~~~~$\omega$ (deg) & \plbOm \\
~~~~$M$\,sin\,$i$ (\mjupe) & \plbMsini \\
~~~~$a$ (AU) & \plbA \\
\sidehead{HD 168443 c}
~~~~$P$ (days) & \plcPer \\
~~~~$T_c\,^{a}$ (JD -- 2,440,000) & \plcTc \\
~~~~$T_p\,^{b}$ (JD -- 2,440,000) & \plcTp \\
~~~~$e$ & \plcEcc \\
~~~~$K$ (m\,s$^{-1}$) & \plcK \\
~~~~$\omega$ (deg) & \plcOm \\
~~~~$M$\,sin\,$i$ (\mjupe) & \plcMsini \\
~~~~$a$ (AU) & \plcA \\
\sidehead{System Properties}
~~~~$\gamma$ (m\,s$^{-1}$) & \plGamma \\
~~~~$\mathrm{d}v/\mathrm{d}t$ (m\,s$^{-1}$\,yr$^{-1}$) & \plTrend \\
\sidehead{Measurements and Model}
~~~~$N_{\mathrm{obs}}$ & \plNobs \\
~~~~rms (\mse) & \plRMS \\
~~~~$\chi^2_{\mathrm{\nu}}$ & \plChiNu \\
[-2ex]
 \enddata
 \tablenotetext{a}{Time of transit.}
 \tablenotetext{b}{Time of periastron passage.}
\end{deluxetable}

\begin{figure}[h]
\includegraphics[width=8.2cm]{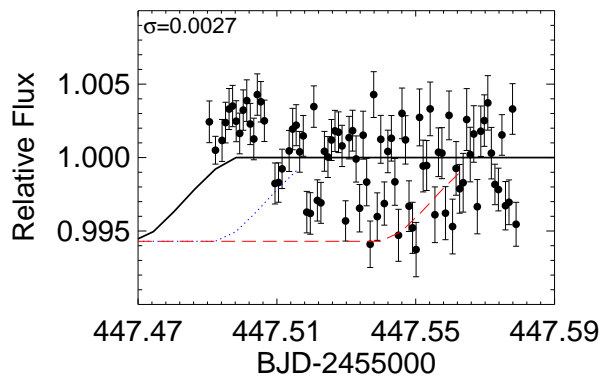}
\caption{Relative photometry from CTIO. The standard deviation of 0.0027 in 
the normalised flux is precise enough to detect the predicted transit, 
which would causes a 0.006 decrease. The solid line represents the predicted 
transit after ephemeris refinement using Keck HIReS data, with the predicted 
egress occurring right before the measurements were acquired. The dotted blue 
and the dashed red lines represent the 1-$\sigma$ and 3-$\sigma$ uncertainties
in the time of egress.}
\label{ctio:phot}
\end{figure}

\begin{figure}[h]
\includegraphics[width=8.2cm]{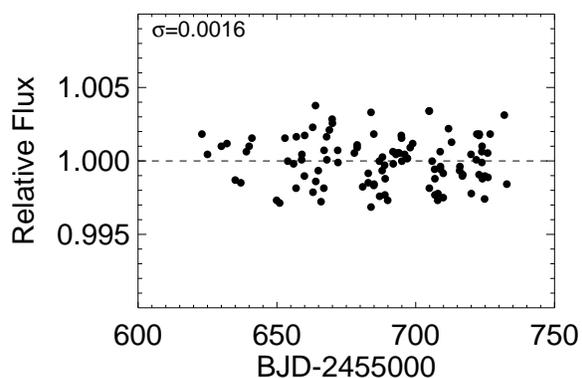}
\caption{Photometry from the T8 APT consisting of 107 measurements over the 
span of 109 days. The dashed line represents the normalised flux. The standard 
deviation from the mean is 0.0016.}
\label{fig:phot_APT}
\end{figure}

\begin{figure}[h]
\includegraphics[width=8.2cm]{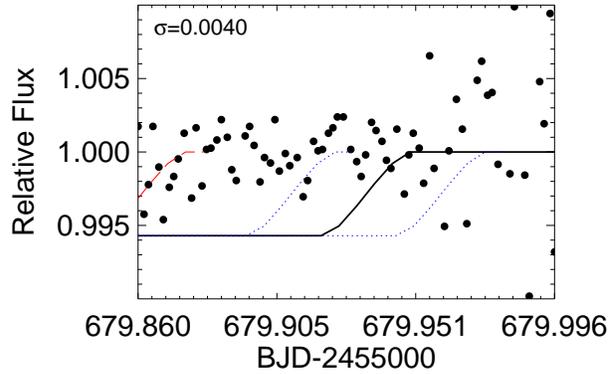}
\caption{Photometric observations of HD~168443 acquired with the T8 APT 
at Fairborn Observatory during the predicted transit of 2011 April 28. The 
standard deviation of 0.0040 flux units is sufficient to detect the predicted 
transit egress (solid line) if present. The dotted blue lines represent the 
$\pm~1-\sigma$ uncertainty in the transit time. There is no evidence for
an egress event.}
\label{fig:APTPHOT1}
\end{figure}

\begin{figure}[h]
\includegraphics[width=8.2cm]{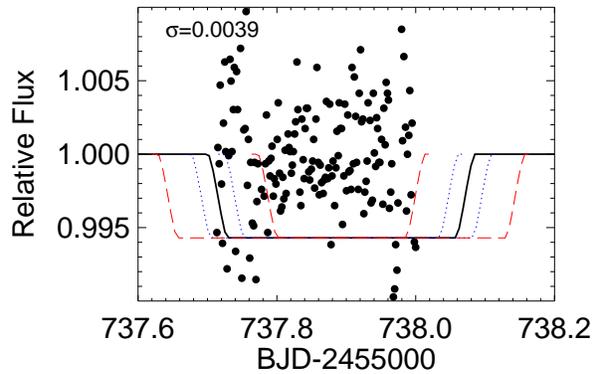}
\caption{Photometric observations of HD~168443 with the T8 APT during 
the predicted transit of 2011 June 25. The standard deviation of 0.0039 
flux units is sufficient to detect the predicted transit egress (solid line) 
if present. The dotted blue lines represent the $\pm~1-\sigma$ uncertainty 
in the transit time. Again, there is no indication of a transit.}
\label{fig:APTPHOT2}
\end{figure}

\begin{figure}[h]
\includegraphics[width=8.2cm]{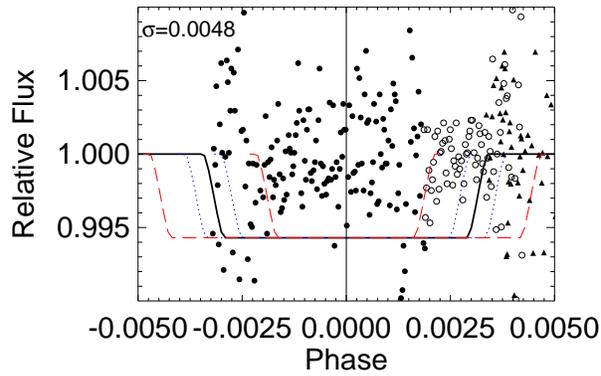}
\caption{Phase diagram of the observations from all three nights of transit 
monitoring. The filled and unfilled circles are the T8 APT measurements 
from 2011 June 25 and April 28, respectively. The filled triangles are the 
CTIO observations from 2010 September 7. As in earlier figures, the solid 
line represents the predicted flux changes during a transit. The dotted blue 
and the dashed red lines represent $\pm~1-\sigma$ and $\pm~3-\sigma$ 
deviations in the time of transit, respectively.}
\label{fig:APTPHOT_phase}
\end{figure}

\end{document}